\documentclass[aps,prb,twocolumn,citeautoscript,superscriptaddress]{revtex4-1}  
\usepackage{amsmath,amssymb,mathrsfs,bm,feynmf,setspace,xspace,soul}  
\usepackage{graphicx}  
\usepackage[tight]{subfigure}    
\usepackage{color} 
\usepackage[colorlinks=true]{hyperref}  
\hypersetup{
    bookmarks=true,         % show bookmarks bar?
    unicode=false,          % non-Latin characters 
    pdftoolbar=true,        % show Acrobat
    pdfmenubar=true,        % show Acrobat 
    pdffitwindow=false,     % window fit to page when opened
    pdfstartview={FitH},    % fits the width of the page to the window
    pdftitle={My title},    % title
    pdfauthor={Author},     % author
    pdfsubject={Subject},   % subject of the document
    pdfcreator={Creator},   % creator of the document
    pdfproducer={Producer}, % producer of the document
    pdfkeywords={keyword1} {key2} {key3}, % list of keywords
    pdfnewwindow=true,      % links in new window
    colorlinks=true,       % false: boxed links; true: colored links
    linkcolor=magenta, %red,          % color of internal links (change box color with linkbordercolor)
    citecolor=blue,        % color of links to bibliography
    filecolor=magenta,      % color of file links
    urlcolor=cyan           % color of external links
} 

\newcommand{\al}{\alpha}

\newcommand{\de}{\delta} 
\newcommand{\e}{\epsilon}

\newcommand{\la}{\lambda}
\newcommand{\La}{\Lambda}

\newcommand{\s}{\sigma}

\newcommand{\w}{\omega}
\newcommand{\W}{\Omega}
\newcommand{\De}{\Delta} 
\newcommand{\G}{\Gamma}

\newcommand{\pd}{\partial}

\newcommand{\beq}{\begin{equation}}
\newcommand{\eeq}{\end{equation}}
\newcommand{\ba}{\begin{array}{ccc}}
\newcommand{\ea}{\end{array}}
\newcommand{\nn}{\nonumber \\}

\def\bea{\begin{eqnarray}}
\def\eea{\end{eqnarray}}
\newcommand{\bml}{\begin{multline}}

\newcommand{\eeqm}{\end{multline}}

\newcommand{\bsp}{\begin{split}}
\newcommand{\esp}{\end{split}}

\renewcommand{\b}[1]{{\bf #1}}
\renewcommand{\t}{\tilde}

\newcommand{\inv}{^{-1}}

\newcommand{\mc}{\mathcal}

\renewcommand{\t}{\tilde}
\newcommand{\ra}{\rightarrow}
\newcommand{\ts}{\thinspace{}}
\newcommand{\req}[1]{Eq.\thinspace(\ref{eq:#1})} 

\newcommand{\rfig}[1]{Fig.\ts\ref{fig:#1}}

\newcommand{\ie}{{\em i.e.\/}\@\xspace}

\DeclareMathOperator{\im}{Im}
\DeclareMathOperator{\re}{Re}

\newcommand{\og}{\mc O_{\! g}} 
\newcommand{\mo}{\mc O} 
\renewcommand{\d}{{\rm d}} 

\begin{document}  

%\title{Quantum critical asymptotics and sum rules}
%\title{Constraining the quantum critical dynamics of the 2+1D Ising model and beyond}
\title{Constraining quantum critical dynamics:
2+1D Ising model and beyond} 
 \author{William Witczak-Krempa} 
 \affiliation{Perimeter Institute for Theoretical Physics, Waterloo, Ontario N2L 2Y5, Canada}
 \date{\today}
\begin{abstract}   
Quantum critical (QC) phase transitions generally lead to the absence of quasiparticles. The resulting
correlated quantum fluid, when thermally excited, displays rich universal dynamics. 
We establish non-perturbative constraints on the linear-response dynamics of conformal QC systems at finite 
temperature, in spatial dimensions above one. 
Specifically, we analyze the large frequency/momentum asymptotics of observables, which we use to derive powerful sum rules
and inequalities. The general 
results are applied to the O($N$) Wilson-Fisher fixed point, describing the QC Ising model when $N=1$.
We focus on the order parameter and scalar susceptibilities, and the dynamical shear viscosity.
% We further illustrate the extension to non-conformal QC points using the quadratic band touching of fermions.
Connections to simulations, experiments and gauge theories are made.  
%Crucial information about quantum critical systems is encoded in the large-frequency behavior of 
%observables at finite temperature.
\end{abstract}
\maketitle   
%\tableofcontents  

The quantum Ising model in two spatial dimensions (2+1D), \emph{e.g.}\ on a square lattice, undergoes a quantum critical 
(QC) phase transition  
as the ratio of the transverse magnetic field to the exchange coupling is tuned. It is the archetypal example of a non-trivial 
2+1D QC point, possibly the simplest one with $Z_2$ symmetry, but lacks an exact solution contrary to its lower dimensional counterpart.     
Rather than having quasiparticles excitations, present in the para/ferromagnetic phases, the spectrum at the QC point
is \emph{continuous}. Various methods such as Monte Carlo simulations\cite{mc-rev}, field theory
expansions\cite{jzj,mc-rev,book}, and recently conformal bootstrap\cite{ising-bootstrap}, have shed light on the critical exponents characterizing its 
thermodynamics and groundstate correlations. 
%Many of the results are in fact derived in the context of the \emph{classical} 3+0D model.
In contrast, little is known about its quantum \emph{dynamical} propertiesx
at finite temperature\cite{book,csy}, which are not only important to understand the nature of this strongly correlated quantum fluid
but also of clear relevance to experiments.    

In this article we study QC dynamics, with a focus on the quantum O$(N)$ Wilson-Fisher fixed point %for $2<D<4$
which describes the QC transition for the quantum Ising ($N\!=\!1$) and XY ($N\!=\!2$) models, and
the N\'eel transition
in certain antiferromagnets ($N\!=\!3$). 
Focusing on a large class of experimentally relevant observables, we establish non-perturbative results for the 
large frequency/momentum asymptotic behavior and sum rules. These provide strong constraints on the universal scaling
functions characterizing the system's low-energy responses. The exact sum rules can be  
seen as generalizations of the celebrated $f$-sum rule to scale invariant systems. Our results provide rigorous means to
assess approximations, constrain numerical results, and ultimately assist with the analysis of experimental data.
The methods we use partly rely on the conformal symmetry of the QC point, present for the O$(N)$ Wilson-Fisher
fixed point. However, the key ideas are more general, and they greatly generalize the recent analysis\cite{ope1} for
the dynamical conductivity of 2+1D conformal field theories (CFTs). The paper is organized as follows: 
We first establish general properties regarding the asymptotics
and sum rules of CFTs, and subsequently apply them to the Wilson-Fisher theory, % we then present
% a non-conformal example (quadratic band touching)
and finally give a broad outlook, including a discussion regarding the implications for Monte Carlo 
simulations.  
\begin{figure}   
\centering
\includegraphics[scale=.46]{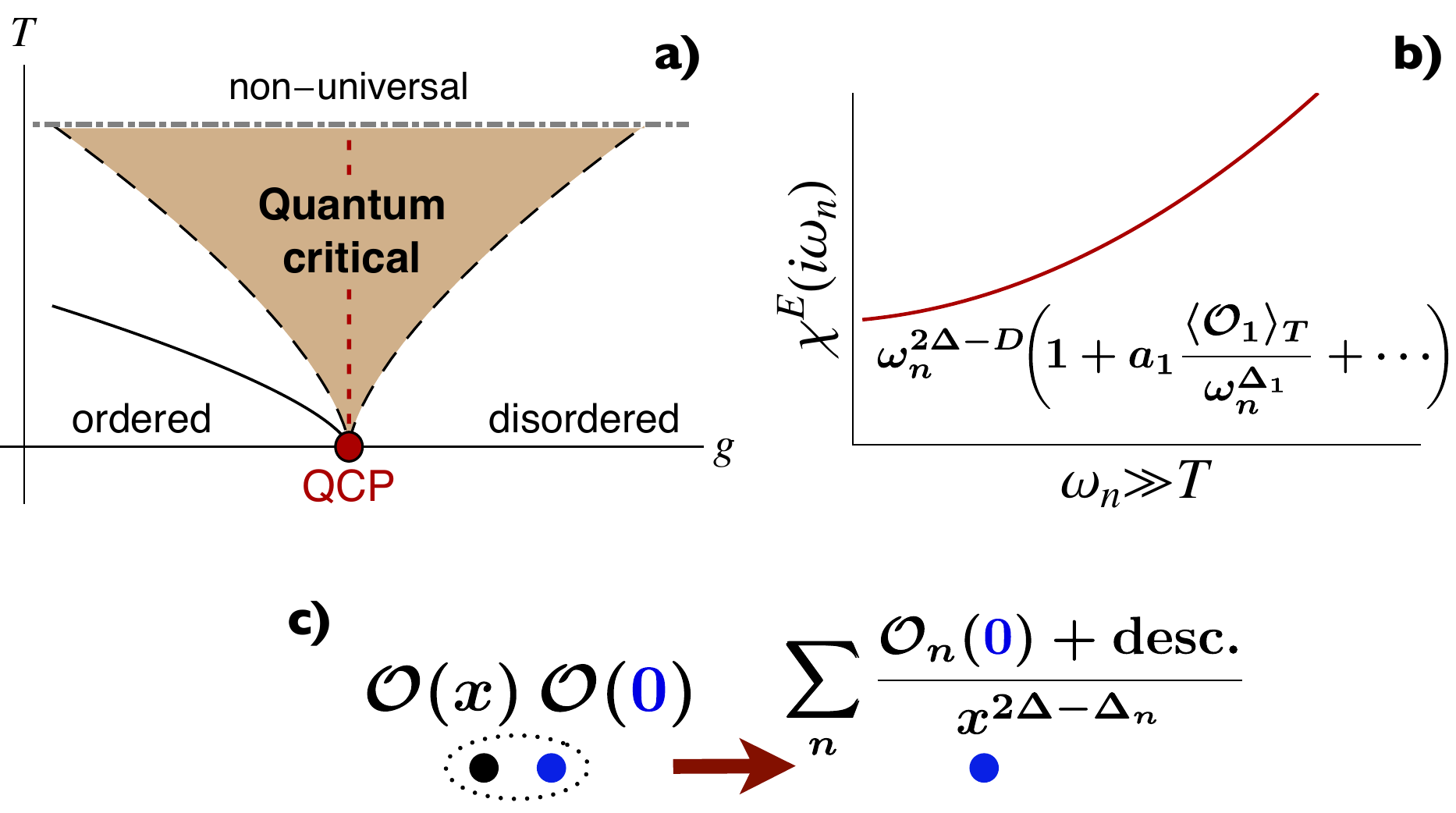}     
\caption{\label{fig:1} {\bf a)} Phase diagram near a quantum critical point (QCP). {\bf b)} Asymptotic 
behavior of the Euclidean susceptibility associated with an operator $\mc O$ of scaling dimension $\De$: 
$\chi^{\rm E}(i\w_n)=\langle \mc O(-\w_n)\mc O(\w_n)\rangle_T$.
{\bf c)} Schematic operator product expansion (OPE) determining the asymptotics of $\chi$. ``desc.'' denotes the descendants 
of the primary $\mc O_n$ (dimension $\De_n$).  
} 
\end{figure}    

% \section{Asymptotics} 
% \label{sec:asympt}   
{\bf Asymptotics and OPE:}    
We consider a thermally excited system tuned to a QC point via a non-thermal parameter $g$. 
In the phase diagram \rfig{1}a, this corresponds to the line in the QC fan at $g\!=\!g_c$ and $T\geq0$.
We are interested in the linear-response dynamics at finite temperature, more precisely in the retarded dynamical susceptibility 
associated with a bosonic observable $\mc O$, such as the energy or charge density:
$\chi^R(t,\b x)=-i\Theta(t)\left\langle \left[\mc O(t,\b x),\mc O(0,\b 0)\right] \right\rangle_T$,
where the average is taken over the thermal ensemble. We set $\hbar=k_B=c=1$; $c$ is the characteristic speed near the QC point.
We will often work in Fourier space:
$\chi^R(\w,\b k)=\int\! \d t\d^d\b x \chi^R(t,\b x)e^{i\w t-i\b k\cdot\b x}$, where $\w$ is the real frequency
and $\b k$ the momentum. Using $T$, the only energy scale available, $\chi^R$ can be rewritten to make its scaling properties manifest: 
\begin{align} \label{eq:scaling}
  \chi^R(\w,\b k) = T^{2\De_\mo-D}\, \Phi\!\left(\frac{\w}{T}, \frac{\b k}{T^{1/z}}\right)\,,
\end{align}
where $\De_\mo$ is the scaling dimension of $\mo$, %, such that $\langle \mo(x)\mo(0)\rangle_{T=0}=1/|x|^{2\De_\mo}$. 
$\Phi$ is a universal scaling 
(complex) function, and $z$ the dynamical critical exponent. This scaling structure emerges at low energies,
\ie\ $\w,|\b k|,T\ll\La_{\rm UV}$, where $\La_{\rm UV}$ is a microscopic lattice energy scale, represented by the horizontal dot-dashed
line in \rfig{1}a. We emphasize that in this regime the ratios $(\w,|\b k|^z)/T$ can be arbitrary.  
We introduce the corresponding universal response function  
\begin{align} \label{eq:def_R}
  \mc R(\w,\b k)=\frac{\chi^R(\w,\b k)}{i\w-0^+}\,, %=  T^{2\De_\mo-D-1} \Phi_{\mc R}\!\left(\frac{\w}{T}, \frac{\b k}{T^{1/z}}\right)\,,
\end{align}
using the Kubo prescription. \emph{E.g.}\ if $\mc O=J_x$ is a conserved current,
$\chi_{xx}^R$ is the $xx$-polarization function and $\mc R_{xx}(\w,\b 0)=\s_{xx}(\w)$ the dynamical conductivity.    
Due to the strong interactions and the resulting absence of quasiparticles in generic QC systems,
little is known about these universal responses, and our goal is to unravel some of their robust properties.
First, let us begin with the large-frequency regime, $\w\gg T,|\b k|$, where the dynamics are near those of the groundstate.   
%of the susceptibility and response function (\rfig{1}b),
These can be elegantly studied via the operator product expansion\cite{wilson} (OPE) of $\mc O$ with itself. 
The OPE is an operator relation and does not depend on temperature.   
%property of the \emph{groundstate}: 
For a general QFT, it is a short time/distance  
expansion that captures the behavior of the operator product $\mc O_1(t,\b x)\mc O_2(0,\b 0)$ as $t,|\b x|\ra 0$, which by locality, 
can be expressed as an infinite sum of operators evaluated at $t,|\b x|=0$. We will mostly
focus on CFTs, which have $z=1$ and describe a large class of experimentally relevant QC phase transitions such as those in the quantum Ising and XY models.
In a CFT, the $\mc O\mc O$ OPE\cite{ferrara,polyakov74} of a primary operator $\mc O$ with scaling dimension $\De_{\mc O}$ reads  
(\rfig{1}c)
% \begin{align}
%   \mc O(x)\mc O(0) \overset{x\ra 0}= \frac{C_{\mc O}}{x^{2\De_{\mc O}}}\Tens_0(x) + \sum_n C_n \frac{\mc O_n(0)}{x^{2\De_{\mc O}-\De_n}} \Tens_n(x)
% \end{align}
\begin{align} \label{eq:ope}
  \mc O(x)\mc O(0) = %\frac{C_{\mc O}(x)}{x^{2\De_{\mc O}}} + 
\sum_{\mc O_n\, {\rm primary}} \!\frac{C_n\big(x,\tfrac{\pd}{\pd y}\big)}{|x|^{2\De_{\mc O}-\De_n}} \mc O_n(y)\Big|_{y=0}\,,
\end{align}
which is expressed in imaginary time $\tau$: $|x|^2=\tau^2+\b x^2$. A primary operator transforms
homogeneously under conformal transformations; \emph{e.g.}\ conserved currents and the order parameter in the O$(N)$ model.
The sum in \req{ope} is over primaries
$\mc O_n$ with scaling dimensions $\De_n$;    
it includes the identity (dimension 0). 
%and $C_n$ is a real number, an OPE coefficient. 
The differential operator $C_n\big(x,\tfrac{\pd}{\pd x}\big)$ is homogeneous under $x\ra b x$, and encodes the contributions from the descendants of $\mc O_n$ (obtained by applying derivatives to $\mc O_n$). 
%and depends only on the dimensionless coordinates, $\hat x_\mu=x_\mu/x$. 
% In Fourier space, 
% \begin{align}
%   \mc O(-k)\mc O(k)= \mc C_{\mc O}(\hat k) k^{2\De_{\mc O}-D} + \sum_n \mc C_n(\hat k) k^{2\De_{\mc O}-\De_n-D} \mc O_n(0)
% \end{align}
Going to Fourier space and taking a thermal expectation value (TEV) we obtain a key result (\rfig{1}b): 
the $|k|\gg T$ behavior of the Euclidean susceptibility, 
\begin{align} \label{eq:gen-asympt}
  \chi^{\rm E}(k) = |k|^{2\De_{\mc O}-D}\!\!\! \sum_{\mc O_n\, {\rm primary}}\!\!\! \left(c_n(k) \frac{\langle \mc O_n\rangle_T}{|k|^{\De_n}}+\dotsb\right)\,,
%&= k^{2\De_{\mc O}-D} \left( c_{\mc O}(\hat k) + \sum_n c_n(\hat k) d_n \left(\frac{T}{k}\right)^{\!\De_n} \right) 
\end{align}
where $|k|^2=\w_n^2+\b k^2$, $\w_n=2\pi Tn$ is a Matsubara frequency. 
The dimensionless functions $c_n$ encode the appropriate $k$-space tensor structure 
(and can contain logarithms).
The dots correspond to higher powers of $T/|k|$ 
arising from the descendants of $\mc O_n$.
Crucially, a scaling operator will acquire a TEV, 
$\langle \mc O_n\rangle_T = d_n\, T^{\De_n}$,
since $T$ is the only energy scale. $d_n$ is a universal real number. Substituting this into \req{gen-asympt} 
we obtain a general expression for the large-$k$ asymptotic expansion of $\chi$.  
We see that the lowest dimension operators appearing in the OPE dictate how the susceptibility  
approaches its groundstate value as $T/|k|\ra 0$. 
To obtain the real quantum dynamics, we can analytically
continue the imaginary frequency expansion \req{gen-asympt} to real frequencies\footnote{We stay away from the lightcone
$\w^2=\b k^2$.} termwise, with
the replacement $i\w_n\!\to \w+i0^+$. This follows from the structure of the OPE 
and the spectral representation connecting the Euclidean and retarded susceptibilities 
(see App.\ts\ref{app:continuation} for
an extension of the proof in [\onlinecite{caron09}]). 

Interestingly, unitarity and conformal symmetry constrain
the scaling dimensions of these operators\cite{mack77}: $\De_n\geq (D-2)/2$. This leads to important inequalities for
dynamical susceptibilities.
Let us work in 2+1D and consider a putative low-energy susceptibility $\chi(\omega)$ 
obtained from an experiment or simulation, and express it as
\begin{align}
  \chi(\w) \overset{\omega\gg T}{=} \omega^{2\de_1-3}\,\bigg[a + b \left( \frac{T}{\omega} \right)^{\!\de_2} +\dotsb\bigg]\,,
\end{align} 
at large frequencies. Finding either $\de_i<1/2$ would violate   
unitarity bounds and thus rule out a conformal QC point\cite{maldacena-priv}.  
For Wilson-Fisher QC points, we shall see that the stronger condition, $\de_2>1.4$, holds.
Before applying the above general results to those 
CFTs, we discuss how the asymptotics can be used to prove
sum rules for any susceptibility (2-point function). 

{\bf Sum rules:}
We put forth a powerful sum rule for the real frequency quantum dynamical response function: %for the spectral function of the 2-point correlator of $\mc O$,
%namely $\im\chi^R(\w,\b k)$:
\begin{align} \label{eq:sr-R}
 \int_{-\infty}^\infty \!\frac{d\w}{\pi} \re\de\mc R(\w,\b k) = -\de\chi^\infty\,.
\end{align}
$\de\mc R$ is defined as in \req{def_R}, with a modified susceptibility $\chi\!\to\!\de\chi$, defined below.
% \begin{align} \label{eq:sr-chi}
%  \int_{-\infty}^\infty \frac{\d\w}{\pi\w} \im \de\chi^R(\w,\b k) = \chi^R(0,\b k)-\de\chi^\infty \,,
% \end{align}
The sum rule is independent of 
small frequency details, and fundamentally relies on the retarded causal structure of $\chi^R$. More precisely it is the zero-frequency 
limit of the Kramers-Kronig transform for the modified susceptibility, $\de\chi^R(\w,\b k)-\de\chi^\infty$, 
which we now discuss. 
In the QC scaling regime, $\chi$ does not usually decay at large frequencies unlike 
on the lattice because it encodes excitations at all scales.
To formulate the sum rule, we thus generally need to subtract terms, denoted by $\t\chi$, from $\chi$ to remove its large-$\w$
divergence\cite{sum-rules,ws,ws2,william2}:  
$\de \chi(\W,\b k)=\chi(\W,\b k)-\t\chi(\W)$, where $\W$
is a complex frequency in the upper half-plane. 
In some cases, one further needs to subtract a remaining constant: 
$\de\chi^\infty = \de\chi^{\rm E}(\W\!\to\!i\infty, \b k)$, %(\W\ra +i\infty,\b k)$,
where the limit is taken at fixed $T$. We emphasize that $\t\chi(\W)$ is
\emph{momentum-independent} because the asymptotic behavior \req{gen-asympt} depends on powers of 
$T/\sqrt{\w_n^2+|\b k|^2}$ due to the asymptotic re-emergence of Lorentz invariance (broken by $T$),
and we fix $\b k$ as we take $\w_n\gg T$.    
% In this equation, $\chi^R(0,\b k)$ 
% from the RHS of \req{sr-chi} does not appear since it has been absorbed in a $\de(\w)$ contribution to $\mc R$.  
We note that the correlation functions studied 
here only depend on the magnitude of the momentum $|\b k|$, which implies that $\im\chi^R$,$\re\mc R$
are $\w$-even functions, so that the integral Eq.\thinspace(\ref{eq:sr-R}) can be written for $\w\geq 0$. %converted to $2\int_0^\infty \d\w$. 

The highly non-trivial and theory dependent information is contained in the subtraction terms $\t\chi,\de\chi^\infty$
which are determined from the large-frequency behavior, \ie from the leading terms in OPE, \req{ope}. 
We now derive some general properties of the subtractions. 
First, the main subtraction $\de\chi=\chi-\t\chi$ is generally required because the leading asymptotic $|k|\gg T$ behavior 
of $\chi^{\rm E}(k)$ is $|k|^{2\De_{\mc O}-D}$, and most operators have $2\De_{\mc O}> D$. %, and the leading term diverges as $|k|\ra\infty$.  
In contrast, in almost all cases the subtraction of a constant is not needed, \ie\ $\de\chi^\infty=0$. Indeed, from
\req{gen-asympt} this constant can be  
non-zero only if the $\mc O\mc O$ OPE contains an operator $\mc O_*$ with dimension $\De_*=2\De_{\mc O}-D$.
Moreover, $\mc O_*$ needs to have a non-zero TEV. In which case, $\de\chi^\infty\propto \langle \mc O_*\rangle_T$ and the constant
of proportionality is the corresponding OPE coefficient. A further necessary condition for $\de\chi^\infty\neq 0$ 
is $\De_{\mc O}\geq (3D-2)/4$
because of the unitarity bound\cite{mack77} on $\De_*$. %$\De_*\geq (D-2)/2$   
A \emph{generic} case where the subtraction $\de\chi^\infty$ appears is %such an operator $\mc O_*$ exists is  
for a 2-point function of $T_{\mu\nu}$, the stress tensor, because the latter has scaling dimension $D$.  
In this case
$\mc O_*\!=\!T_{\mu\nu}$ since the stress tensor generally appears in the $T_{\mu\nu}T_{\la\varepsilon}$ OPE. Below we will 
%see an example where $\de\chi^\infty$ is finite 
the consequences of this for the shear viscosity. %for a correlator of $T_{xy}$.   

{\bf O$(N)$ model:} 
We now apply the above general results to the QC point of the quantum O$(N)$ model\cite{polyakov-book,book}
in dimensions $2<D<4$. This is the famous Wilson-Fisher conformal fixed point. It describes  
a variety of experimentally relevant quantum phase transitions: Ising ($N\!=\!1$), XY ($N\!=\!2$), etc.
An exact solution exists at $N\!=\!\infty$, which we will use to perform non-trivial checks.
As a field theory, the O$(N)$ (non-linear sigma) model is defined by the action $S=\int\! \d^D\!x \frac{1}{g}\pd_\mu\varphi_a\pd_\mu\varphi_a$, where $\varphi_a(x)$ is a 
real $N$-component vector field of fixed norm $\varphi_a\varphi_a=1$. 
As the coupling $g$ is increased the system undergoes a  
QC phase transition at $g=g_c$ from a broken symmetry phase to a symmetric one for $g>g_c$ (\rfig{1}a).  

For our asymptotics/sum rule analysis we need the list of operators $(\mc O_n,\De_n)$ with low dimensions $\De_n\leq D$. 
These are known from large-$N$ and small $(4-D)$ expansions\cite{jzj,mc-rev}, Monte Carlo\cite{mc-rev}, 
non-perturbative bootstrap\cite{ising-bootstrap,boot-vector,boot-frac}, etc. 
The first one being the order parameter field $\phi_a$ with dimension $\De_\phi=(D-2+\eta_\phi)/2$, %$(D-2)/2<\De_\phi<1$ 
where $\eta_\phi$ is the field's anomalous dimension. 
The following O$(N)$-invariant operators will also appear: %our analysis are %the fundamental scalar $(\phi_a,\De_\phi)$, 
the ``thermal'' operator $(\og,\De_g)$,  
the conserved currents $(J^\mu_{ab},D-1)$, and the stress tensor $(T_{\mu\nu},D)$. The dimensions of the currents and stress tensor
receive no anomalous corrections because they are protected by symmetries.
The operator $\og$ (often denoted by $\varepsilon$ in the context of the Ising model) is associated with
the Lagrange multiplier field $\la(x)$ that constrains $\varphi_a\varphi_a=1$ in the O$(N)$ model. It has
dimension $\De_g=D-1/\nu$, where $\nu$ is the correlation length exponent; 
for the $D\!=\!3$ Ising case\cite{mc-rev,ising-bootstrap}, $\De_g=1.413$.    
It is directly related to the singlet $\phi^2$,
and tunes the system away from the QC point. Being the only relevant O$(N)$-symmetric scalar, 
it is the most important operator as it dominates the asymptotic quantum dynamics:
we will see that it generally gives the first finite-$T$ correction. %subleading term in the large-$k$ asymptotic expansion. 
This was recently shown\cite{ope1} to be the case for the conductivity of the O$(N)$ model, and observed numerically\cite{ope1} for $N=2$. 
Given the generality of our OPE analysis, we infer that this ``dominance'' of the
relevant symmetric scalar is a generic property of QC transitions.  

 {\bf Order parameter susceptibility:}  
We first study $\chi_{ab}(k)=\langle\phi_a(-k) \phi_b(k)\rangle_T$, \ie the order parameter susceptibility.  
It is one of the simplest observables, and yields the low-energy staggered spin susceptibility 
of quantum antiferromagnets with transitions in the O$(N)$ universality class. We begin by analyzing its asymptotics.  
By symmetry, and from the knowledge of the operators with low dimensions we can write the leading terms in the $\phi_a\phi_b$ OPE:
% \begin{multline}
%   \phi_a(x)\phi_b(0) = \frac{\de_{ab}}{x^{2\De_\phi}} + \frac{C_{\!\phi\phi g}\de_{ab}\og(0)}{x^{2\De_\phi-\De_g}} \\
%   %+ \frac{C_{\!\phi\phi J} x_\mu J^\mu_{ab}(0)}{x^{2\De_\phi -(D+1)}} 
% + \frac{C_{\!\phi\phi T} x_\mu x_\nu T_{\mu\nu}(0)}{x^{2\De_\phi -D+2}}+\dotsb
% \end{multline} 
\begin{align} \label{eq:ope-phi}
 \phi_1(x)\phi_1(0)\! =\! \frac{C_\phi}{x^{2\De_\phi}} + \frac{C_{\!\phi\phi g}\og(0)}{x^{2\De_\phi-\De_g}} 
  %+ \frac{C_{\!\phi\phi J} x_\mu J^\mu_{ab}(0)}{x^{2\De_\phi -(D+1)}} 
+ \frac{C_{\!\phi\phi T} x_\mu x_\nu T_{\mu\nu}(0)}{x^{2\De_\phi -D+2}}+\dotsb
\end{align} 
where we focus on $a,b=1$ since $\chi_{ab}$ is diagonal by virtue of O$(N)$ symmetry. 
We have omitted the contribution from the currents $J^\mu_{ab}$ because they have vanishing TEV
(no excess charge or net current in the thermal ensemble).  
Taking the TEV of \req{ope-phi} gives the asymptotic behavior
\begin{multline}  \label{eq:asy-phi}
    \chi_{11}^{\rm E}(i\w_n,\b k) = |k|^{2\De_{\phi}-D} \bigg[ \mc C_\phi + \mc C_{\phi\phi g}d_{g} \left|\frac{T}{k}\right|^{\De_g} \\
      + \mc C_{\phi\phi T} \frac{k_\mu k_\nu}{k^2} d_{T}^{\mu\nu} \left|\frac{T}{k}\right|^D+\dotsb \bigg]\,.
\end{multline}
$C_\#/\mc C_\#$ in Eqs.\ts(\ref{eq:ope-phi})/(\ref{eq:asy-phi}) are real OPE coefficients in position/momentum space,
which can be obtained from groundstate 3-point functions.  
As anticipated, the first subleading term comes from the relevant scalar $\og$. The next term arises from the
stress tensor, where $\langle T_{\mu\nu}\rangle_T=d_T^{\mu\nu}T^D$ is diagonal. 
At $N=\infty$, $\De_\phi=(D-2)/2$ saturates the unitarity bound, but finite $N$ fluctuations lead to
a small anomalous dimension $\eta_\phi\ll 1$\cite{jzj,mc-rev,ising-bootstrap,boot-vector}. 
%as has been shown using Monte Carlo, large-$N$ and small $(4-D)$ expansions, and conformal bootstrap. 
The OPE coefficients $\mc C_{\phi\phi g},\mc C_{\phi\phi T}$ are generally finite 
and can be computed using a $1/N$ expansion for instance. $C_{\phi\phi g}$ has been computed using
bootstrap\cite{el-showk2014} and
Monte Carlo\cite{caselle15} for $N\!\!=\!1$.
From the above expansion, we 
can derive the sum rule for $\chi_{ab}$. First, for any $N$, $\chi_{ab}$ decays sufficiently fast at 
large frequencies so that the subtractions vanish, $\t\chi_{ab}=\de\chi_{ab}^\infty=0$, and the sum rule takes its simplest form:  
\begin{align} \label{eq:sr-phi}
  \int_0^\infty \!\d\w \re \mc R_{ab}(\w,\b k) =0\,, 
\end{align}
where $\mc R_{ab}(\w,\b k) = \chi_{ab}^R(\w,\b k)/(i\w-0^+)$ is the response. 

When $N=\infty$, we have the exact solution for $2<D<4$: $\chi_{ab}^{\rm E}(i\w_n,\b k)=\de_{ab}/(\w_n^2+\b k^2+m_T^2)$, where $m_T=\Theta_d T$ 
is the thermal mass, and $\Theta_d$ is a positive number\cite{petkou98}, see App.\ts\ref{app:scalar}.  
Expanding for $|k|\gg T$, we get  $\chi_{ab}^{\rm E}(k)= \tfrac{1}{k^2}\big[1- (\tfrac{m_{^T}}{k})^2 + (\tfrac{m_{^T}}{k})^4+\dotsb\!\big]$.
% \begin{align}
%   \chi_{ab}(i\w_n,\b k) %= \frac{1}{k^2}\sum_{l=0}\left(-\frac{m_T^2}{k^2}\right)^l
%   = \frac{1}{k^2}\left(1- \frac{m_T^2}{k^2} + \frac{m_T^4}{k^4}+\dotsb\right) %O\left(\frac{m_T}{k}\right)^6 \right) 
% \end{align}
In agreement with the OPE, \req{ope-phi}, the subleading term $-m_T^2/k^4$ has 
$\De_g^{\!N=\infty}\!=2$ (\ie $1/\nu\!=\!\!D-2$) and is proportional to 
$\langle \og\rangle_T= \sqrt{N} m_T^2$. %=\sqrt{N}\Theta_{\!D}^2T^2$. 
This later TEV is evaluated\cite{ope1} in the $N\!=\!\infty$ limit.  
We note the absence of a contribution from the stress tensor, $\sim m_T^3/|k|^5$. Although the real-space OPE coefficient  
$C_{\phi\phi T}$ in \req{ope-phi} is non-zero, upon Fourier transforming to $k$-space, that term does not contribute to the large-$k$
behavior. This is an artefact of $N=\infty$, where $\phi$ has no anomalous dimension. %For finite $N$, the stress tensor is expected to contribute.
Finally, the sum rule \req{sr-phi} can be easily checked as the spectral function is a sum of (quasiparticle) delta functions.    

{\bf Scalar susceptibility}: 
The scalar susceptibility $\chi_s$ is the 2-point function of the ``thermal'' operator, $\langle \og(-k)\og(k)\rangle_T$. 
It has recently been the focus of attention in the study of the amplitude ``Higgs'' 
mode\cite{endres,pollet12,podolsky12,gazit13,rancon14,tenenbaum14}.  
%which corresponds to the 2-point function of $\varphi^2$.  
Again, we first examine the $\og\og$ OPE. The terms relevant here are given, \emph{mutatis mutandis}, by \req{ope-phi}.
% \begin{align}
%   \og(x)\og(0)= \frac{C_{g}}{x^{2\De_g}} + \frac{C_{\! ggg}\,\og(0)}{x^{\De_g}}
%    + \frac{C_{\! ggT} \,x_\mu x_\nu T_{\mu\nu}(0)}{x^{2\De_g -D+2}}+\dotsb 
% \end{align}
This then leads to the large-$k$ expansion \req{asy-phi} with $(\phi_a,\De_\phi)$ replaced by $(\og,\De_g)$.  
% \begin{align}
%     \chi_s(k) = k^{2\De_g-D} \left( 1 + \mc C_{ggg}d_{\og} \left(\frac{T}{k}\right)^{\!\De_g} 
%       + \mc C_{ggT}\frac{k_\mu k_\nu}{k^2} d_{T}^{\mu\nu} \left(\frac{T}{k}\right)^{\!D}+\dotsb \right)
% \end{align}
With this data, we can derive the sum rule for $\chi_s$. First, $\de\chi_s^\infty=0$ since there is no O$(N)$-singlet with 
dimension $\De_*=2\De_g-D$ in the    
spectrum. 
%The only candidate is $\og$ but being relevant ($\De_g<D$), 
%it cannot have dimension $2\De_g-D$. 
The other ingredient needed to build the sum rule is the term removing the large-$\w$
divergence, $\t\chi_s$. In this case, it is simply the groundstate value of $\chi_s$ at $\b k\!=\!\b 0$:  
$\t\chi_s(\W)=\chi_s^{T=0}(\W,\b 0)=\mc C_g \W^{2\De_g-D}$. 
The sum rule reads:
\begin{align} \label{eq:chi_s-sr}
  %2\int_0^\infty \frac{d\w}{\pi} \frac{\im\left[\chi_s^R(\w,\b k)-\chi_s^R(\w,\b 0)\big|_{T=0}\right]}{\w} &= \chi^R(0,\b k) \\
  \int_0^\infty\! \d\w \re[\mc R_s(\w,\b k) - \mc R_s^{T=0}(\w,\b 0)] = 0\,.
\end{align}
% The sum rule could be used to check the consistency of the claim that the Higgs amplitude mode,
% as seen in $\chi_s$, survives at the
% QC point of the O(2) model.
We can again carry out the asymptotic analysis exactly for $N\!\to\!\!\infty$.  
% , where\cite{book} $\chi_s = -N/\Pi$, with the polarization function $\Pi^{\rm E}(i\w_n,\b k)=  
% T\sum_{\nu_n} \int \frac{d^d\b q}{(2\pi)^d} \chi_{11}^{\rm E}(i\w_n+ i\nu_n,\b k +\b q) \chi_{11}^{\rm E}(i\nu_n,\b q)$; 
% $\chi_{ab}^{\rm E}$ is the order parameter susceptibility studied above. 
%We extract the asymptotic behavior of the susceptibility
The result is (App.~\ts\ref{app:scalar}):
$\chi_s^{\rm E}(k) = -\frac{N}{a_{0}}\, |k|^{4-D} ( 1-a_{g}|\tfrac{T}{k}|^2 -a_{T} |\frac{T}{k}|^D +\dotsb )$,
% \begin{align}
%   \chi_s(i\w_n,\b 0) = -\frac{N}{a_{0}}\, \w_n^{3-d}\left( 1-a_{g}\left(\frac{m_T}{\w_n}\right)^2 -a_{T} \left(\frac{m_T}{\w_n}\right)^{d+1}+\dotsb \right)
% \end{align}
where $a_{\#}$ are $D$-dependent constants.  
%given in App.\ts\ref{app:scalar}. 
Interestingly, the coefficient of the 
subleading term, $a_{g}$, vanishes exactly for $D=3$. This comes from the somewhat surprising fact that the thermal operator $\og$ does not appear
by itself in the $\og\og$ OPE when $D=3$ in the $N\!=\!\infty$ limit. In other words, the $C_{\!ggg}^{D=3}$ OPE coefficient vanishes. 
This does not happen for $D\neq 3$, and we do not expect it to hold at finite $N$ 
in $D=3$. 
Indeed, for the Ising case this coefficient was recently computed using Monte Carlo methods and found to
be finite\cite{caselle15}.
%It would be interesting to verify this with a $1/N$ expansion, 
Finally, the sum rule \req{chi_s-sr} can be checked numerically at $N\!=\!\infty$ (App.\ts\ref{app:scalar}).   
%$\re\mc R_s$ is plotted in \rfig{R_s}. 

{\bf Dynamical shear viscosity:}
Finally we examine a correlator involving the stress tensor. Not only is this of fundamental interest
because it can be defined for any CFT, but it will also reveal the full complexity  
of the sum rule.
We consider the dynamical shear viscosity, $\eta(\w,\b k)=\chi^R_{\eta}(\w,\b k)/(i\w-0^+)$,
obtained from the $T_{xy}$ 2-point function, $\chi_\eta^R$. $T_{xy}$ measures the flux of $x$-momentum
in the $y$-direction, and $\eta$ probes the system's resistance against momentum gradients.  
% \begin{align}
%   \eta(\w,\b k) = -i\frac{\chi_\eta^R(\w,\b k)}{\w+i0^+}=-\frac{1}{\w+i0^+}\int \d^Dx e^{-ik\cdot x}\Theta(t)\langle [T_{xy}(x), T_{xy}(0)]\rangle_T
% \end{align}
% which follows from the 2-point function of $T_{xy}$ via the Kubo formula.
%(The static shear viscosity reads $\eta_0=\lim_{\w\ra0}\lim_{\b k\ra\b 0}\eta(\w,\b k)$.)
The asymptotic behavior of $\eta$ follows from the $T_{xy}T_{xy}$ OPE, which we here formulate in momentum space:
\begin{multline}
  \lim_{|k|\gg |p|}
T_{xy}(k) T_{xy}(-k+p) = \mc C_T |k|^D\de(p) \\ +\mc C_{TTg}|k|^{D-\De_g}\og(p) 
  + C_{TTT}^{\mu\nu}T_{\mu\nu}(p)+\dotsb\,,
\end{multline} 
where here $\b k=0$ for simplicity. 
% \sum_{\mu,\nu}\mc C_{TTT}^{\{\mu\nu \}}
% Upon taking the TEV, this leads to: 
% \begin{align*}
%     \!\!\!\!\!\chi^{\rm E}_\eta(k) \!=\! |k|^D\! \bigg[ \mc C_T + \mc C_{TTg}d_{g}\!\left|\frac{T}{k}\right|^{\De_g} 
%       \!\!\!\!+ \mc C_{TTT}^{\mu\nu}d_{T}^{\mu\nu}\! \left|\frac{T}{k}\right|^{D}\!\!\!\!+\!\dotsb\! \bigg].
% \end{align*}
This can then be used to derive a sum rule for $\eta$, which is more involved than for the response functions considered above.
For one, $\de\chi_\eta^\infty=\! \mc C_{TTT}^{\mu\nu}\langle T_{\mu\nu}\rangle_T$ is non-zero, as was explained above on general grounds for 
2-point functions involving $T_{\mu\nu}$. Second,  
the subtraction involved in $\de\chi_\eta$ is \emph{temperature dependent} because $\og$ is relevant. 
%has $\De_g<D$ for all $N$. 
This leads to the following sum rule for the shear response: 
\begin{align} \label{eq:eta-sr}
  \int_0^\infty\!\!\! \d\w\re\!\big[\eta(\w,\b k) - \mc C_T\w^{d} - \mc A (\w/i)^{d-\De_g}\big] 
  = c_\eta P\,,
\end{align} 
where $d=D-1$, $\mc A=\mc C_{TTg}\langle \og\rangle_T$, $P=\langle T_{xx}\rangle_T$ is the pressure of the CFT,
and $c_\eta=-\pi \de\chi_\eta^\infty/(2P)$
%$c_\eta=-\pi \mc C_{TTT}^{\mu\nu}\langle T_{\mu\nu}\rangle_T/(2P)$
is a dimensionless constant. The second term in the integrand is $\eta^{T=0}(\w,\b 0)$
and mirrors the subtraction in the scalar sum rule. The third one
depends on temperature via $\mc A\propto T^{\De_g}$ and
scales with a non-trivial $\w$ power depending on the correlation length exponent $\nu$ via $\De_g=D-1/\nu$.  
Some QC theories are simpler in that they lack a relevant scalar that condenses at $T>0$, as we now discuss. 

We contrast the above shear sum rule with the simpler ones obtained\cite{sum-rules_son,caron09} for $\mc N=4$ super  
Yang-Mills and pure Yang-Mills, which are gauge theories in $D\!=\!4$. In those cases, the result is as in \req{eta-sr} except 
that the third term in the integrand is absent. This stems from the fact that those theories do not
contain a symmetric relevant scalar like $\og$, \ie they are not realized by fine tuning a symmetric ``mass'' term. 
The massless version of QED in $D=3$ with many Dirac fermions coupled to a U(1) gauge field also satisfies this property,
being a stable \emph{phase}. It will thus have a shear sum rule of the same form as super Yang-Mills.
Finally, we note that shear sum rules analogous to \req{eta-sr} were derived in the context of strongly interacting 
ultracold Fermi gases\cite{taylor10,enss10,goldberger11}, which generally do not have emergent Lorentz symmetry.

{\bf \emph{Outlook}}: 
Our non-perturbative results, via the operator product expansion (OPE), for the asymptotics and sum rules apply to a wide class of conformal
QC points, many of which describe experimentally relevant systems.
It will be interesting to apply the program 
described in this article to theories other than to the O$(N)$ Wilson-Fisher fixed point, treated here, 
or even to non-conformal QC systems.
The strong constraints we have derived will also be useful for the analysis of  
numerical and experimental data. For instance, quantum Monte Carlo is a powerful tool to study QC dynamics in \emph{imaginary} 
time\cite{smakov,krempa_nature,ope1,pollet,swanson,gazit14}, and can be used to study the asymptotic regime where the OPE analysis applies, as was recently shown\cite{ope1} for the conductivity. 
The asymptotics and sum rules will also help with the difficult task of analytically continuing the imaginary time data 
to real time by constraining the allowed scaling functions. Along those lines, our results can be used 
with a novel method\cite{krempa_nature,ope1} of analytic continuation based on the AdS/CFT 
holographic principle\cite{Maldacena}: 
Specific data about a QC
theory can be encoded in holographic physically-motivated Ansatzes for the scaling functions.  
These can then be used to perform the continuation.  

\emph{Acknowledgments}: I am indebted to E.~Katz, S.~Sachdev, and E.S.~S\o rensen for their collaboration on related topics. 
I further acknowledge stimulating exchanges with D.J.~Gross, C.~Herzog, J.~Maldacena, R.~Myers and D.T.~Son. 
Research at Perimeter Institute is supported by the Government of Canada through Industry Canada and by the  
Province of Ontario through the Ministry of Research and Innovation. \\[5.6cm]

%%%%%%%%%%%%%%%%%%%%%%%%%%%%%%%%%%%
%\clearpage
\onecolumngrid  
\appendix 

\begin{center} 
{\Large\bf Supplementary Information} 
\end{center} 

\tableofcontents
\section{Analytic continuation of asymptotic susceptibility}
\label{app:continuation}  
This appendix explains how the quantum critical (QC) dynamics in the near groundstate regime
follow from the operator product expansion (OPE) in imaginary time.
More precisely, we show that the asymptotics at large \emph{real} frequencies, $\w\gg T$, can be obtained from 
the Euclidean frequency result via term by term analytic continuation.
To do so, we adapt the line of reasoning put forth in Ref.\ts\onlinecite{caron09}.
The analysis begins with the spectral representation for the Euclidean susceptibility:  
\begin{align} \label{eq:spectral}
  \chi^{\rm E}(i\W) = \int_{-\infty}^\infty \frac{\d\w'}{\pi} \frac{\rho(\w')}{\w'-i\W}\,,
\end{align}
where $\rho(\w)=\im\chi^R(\w+i0^+)$ is the spectral density of the retarded 2-point function. We have dropped the momentum dependence of $\rho$ 
and any other indices as these are not crucial to the discussion.
Here $\W>0$ so that $i\W$ is a positive frequency along the
imaginary axis, so that taking a principal value of the integral is not necessary. 
At finite temperature we should strictly speaking use Matsubara frequencies, $\W=\w_n=2\pi T n$, in \req{spectral}. However,
it is more convenient to employ the spectral representation to analytically continue the susceptibility to arbitrary frequencies
along the imaginary axis, and more generally to the upper half of the complex plane.   
In writing the above equation, we have assumed that $\rho(\w)$ vanishes 
as $|\w|\to\infty$. If this fails one needs to use a subtracted density, $\de\rho=\im\de\chi^R$, as
was discussed in the main text. Our argument is independent of this complication, on which we shall 
comment towards the end of the section.  

For the cases of interest, namely 2-point functions of bosonic operators, $\rho(\w)$ is $\w$-odd allowing us to reduce 
the integral to positive frequencies:
\begin{align} \label{eq:spectral2}
  \chi^{\rm E}(i\W)=\int_0^\infty\frac{\d\w'}{\pi} \frac{2\w'}{\w'^2+\W^2}\rho(\w')\,.
\end{align}
We use the convention according to which $\rho(\w)$ is positive for $\w>0$.
Now, assuming that $\rho(\w)$ has a large-frequency expansion in powers of $1/\w$, we will show that it leads
to a corresponding expansion in powers of $1/\W$ for $\chi^{\rm E}(i\W)$. Logarithms can appear but do not spoil the correspondence, see Sec.\ts\ref{app:even}. Let us assume that $\rho(\w)$ is bounded by the tail $1/\w^\e$, $0<\e<1$, at large frequencies.
Using the positivity of \req{spectral2}, one can show that $\chi^{\rm E}$ is bounded by $1/\W^\e$
as $\W\to\infty$, and thus vanishes in that limit. It can also be shown 
that the first derivative of \req{spectral2} with respect to $1/\W$ diverges as $\W\to\infty$.
This establishes that $\chi^{\rm E}\to 1/\W^{\bar\e}$ at large $\W$, with an exponent $0<\bar\e\leq \e$.  
The procedure can be adapted to a more general spectral tail, $1/\w^{\ell+\e}$, where $\ell$ is a positive
integer. In this case, one needs to take $\ell+1$ derivatives with respect to $1/\W$. 
The argument can be further iterated for all the terms in the power law expansion of $\rho$ by considering
a modified density $\t\rho$ from which one has subtracted
terms up to the one that is targeted. We have thus established that an expansion for $\rho$ in powers of $1/\w$
leads to a corresponding expansion in powers for $\chi^{\rm E}$. We now turn to the crux of the proof and
show that these expansions are precisely related by termwise analytic continuation, $i\W\leftrightarrow \w+i0^+$.

Let us assume that $\rho(\w)$ contains the power law $1/\w^\al$ for $\w\gg\w_0$, where we are mainly interested in 
the case where the infrared scale is temperature, $\w_0=T$. Consider $\al>0$ but not an even integer, then using \req{spectral2} we get
\begin{align}
  \int_{\w_0}^\infty \frac{\d\w'}{\pi} \frac{2\w'}{\w'^2+\W^2}\,\frac{1}{\w'^\al} = \frac{1}{\W^\al\sin(\pi\al/2)}
  +\frac{2}{\pi}\sum_{n=1} a_n\frac{\w_0^{2n-\al}}{\W^{2n}}\,,
\end{align}
where $a_n= (-1)^{n+1}/(\al-2n)$. We see that the first term is the expected power law, and is independent of 
the infrared scale $\w_0$ defining the asymptotic regime. The remaining terms depend on $\w_0$ and form a series that contains only \emph{even}
powers of $1/\W$. Due to their incompatible powers, these do not interfere with the first term. In addition, the factor 
of $1/\sin(\pi\al/2)$ is exactly as expected from the analytic continuation:
\begin{align} \label{eq:continuation}
  \W^{-\al}\leftrightarrow (-i\w+0^+)^{-\al}=\w^{-\al}[\cos(\pi\al/2)+i\sin(\pi\al/2)]\,.
\end{align}
We can also recover the real part of $\chi^R$ by making use of the Kramers-Kronig transform (an application of the
spectral representation),
\begin{align}
  \re\chi^R(\w+i0^+) = \mc P \!\!\int_0^\infty \frac{\d\w'}{\pi} \frac{2\w'\rho(\w')}{\w'^2-\w^2}\,,
\end{align}
where we have again made use of the fact that $\rho(\w)$ is odd. $\mc P\!\!\int$ denotes the principal value
of the integral.
Using an infrared cutoff $\w_0$, we find that the real part corresponding to a spectral power law tail $1/\w^\al$ is
\begin{align}
  \re\chi^R(\w+i0^+) &\to \mc P \!\!\int_{\w_0}^\infty \frac{\d\w'}{\pi} \frac{2\w'}{\w'^2-\w^2} \frac{1}{\w'^\al} \\
  &= \frac{\cot(\pi\al/2)}{\w^\al} - \frac{2}{\pi}\sum_{n=1} b_n\frac{\w_0^{2n-\al}}{\w^{2n}}
\end{align}
where $\cot x=\cos(x)/\sin(x)$, and $b_n=1/(\al-2n)$. The first term is independent of $\w_0$
and precisely yields the answer expected from \req{continuation},
whereas the remaining series again decouples because it only contains even powers of $1/\w$.  

In the above, we have assumed that $\rho(\w)$ decays to
zero as $\w\to\infty$. If this is not the case, one can analytically continue the terms with positive powers
of $\w$ using \req{continuation}, and apply the above procedure to the modified susceptibility, $\chi-\t\chi-\de\chi^\infty$,
which vanishes as $\w\to\infty$. We refer the reader to the main text for a detailed discussion regarding the subtractions $\t\chi,\de\chi^\infty$.
We have thus shown that the asymptotic expansions of the retarded and Euclidean susceptibilities
are precisely related by a termwise analytic continuation. 

Our results can be explicitly checked for the infrared fixed point of the O$(N)$ model in the $N\to\infty$ limit,
as is discussed in the main text and in the next appendix. In the context of the charge conductivity in 2+1D, this was done
for the O$(N\to\infty)$ model and the Dirac CFT in Ref.\ts\onlinecite{ope1}, and for a wide class of 
conformal QC theories without quasiparticles using AdS/CFT in Refs.\ts\onlinecite{ope1,william2,sum-rules}. 

\subsection{Even powers in the spectral function} \label{app:even}
We now turn to the more subtle case of even powers in $\rho(\w)$. From the basic formula for 
analytic continuation, \req{continuation}, we see that even powers in imaginary frequencies contribute
only to the real part of $\chi^R$. \emph{E.g.}\ $1/\W^2\to -1/\w^2$, which does not contribute to $\rho=\im\chi^R$.
To understand what asymptotic Euclidean term can give rise to a $1/\w^2$ scaling for $\rho$, we
turn to the spectral representation, \req{spectral2},
\begin{align}
  \int_{\w_0}^\infty \frac{\d\w'}{\pi} \frac{2\w'}{\w'^2+\W^2}\,\frac{1}{\w'^2} &= \frac{1}{\pi\W^2}\ln\left(1+\frac{\W^2}{\w_0^2} \right)\,; \\
 &\overset{\W\gg \w_0}{=} \frac{\ln\left(\frac{\W^2}{\w_0^2} \right)}{\pi \W^2}  
  +\frac{1}{\pi}\sum_{n=2} r_n\frac{\w_0^{2n-2}}{\W^{2n}}\,,
\end{align}
where $\w_0$ is the infrared cutoff we have employed above, and $r_n=(-1)^{n}/(n-1)$. The first term decouples from the series due 
to the presence of the logarithm; also note that there is no $1/\W^2$ term in the series.
We thus see that a logarithm in $\chi^{\rm E}$ gives rise to a $1/\w^2$ term in $\rho$. This is fully consistent with the direct analytic continuation
of the term with the logarithm:
\begin{align}
  \frac{\ln\left(\frac{\W^2}{\w_0^2} \right)}{\pi \W^2}\; \xrightarrow{i\W\to \w+i0^+}\; - \frac{\ln\left(\frac{\w^2}{\w_0^2}\right)}{\pi \w^2}
+ \frac{i}{\w^2}\,,
\end{align}
for $\w>0$.
Analogous results hold for higher powers of $1/\w$, \emph{e.g.}\ $1/\w^4$ obtains from 
$(\pi \W^4)\inv\ln(\W^2/\w_0^2)$, etc. We note that such logarithms do not appear in the leading expansion of the response functions of
the Wilson-Fisher QC points studied in this paper.

\section{Scalar susceptibility of the O$(N)$ model at large $N$}
\label{app:scalar}
We give details regarding the asymptotics and sum rule for the scalar susceptibility $\chi_s(k)=\langle \og(-k)\og(k)\rangle_T$
of the O$(N$) Wilson-Fisher fixed point in the large $N$ limit.
Using the O$(N)$ non-linear sigma model defined in the main text, it can be shown\cite{book} that for $N\to\infty$ the
susceptibility is given by 
\begin{align}
  \chi_s(k) = -\frac{N}{\Pi(k)}\,,  
\end{align}
with the scalar polarization function  
\begin{align}
\Pi^{\rm E}(i\w_n,\b k)=  
T\sum_{\nu_n} \int \frac{\d^d\b q}{(2\pi)^d} \chi_{11}^{\rm E}(i\w_n+ i\nu_n,\b k +\b q) \chi_{11}^{\rm E}(i\nu_n,\b q)\,. 
\end{align}
The sum/integral involves
\begin{align} \label{eq:propagator}
  \chi_{ab}^{\rm E}(k)=  \frac{\de_{ab}}{\w_n^2+\b k^2+m_T^2}\,,
\end{align}
the order parameter susceptibility (propagator).
At zero external spatial momentum, we find
\begin{align}
  \Pi^{\rm E}(i\w_n,\b 0) &= \int\! \frac{\d^d\b k}{(2\pi)^d} \frac{1+2n_B(\e_k)}{\e_k}\frac{1}{4\e_k^2+\w_n^2}\,;\\
  &= K_d\int_{m_T}^\infty \!\d\e\, (\e^2-m_T^2)^{(d-2)/2}\; \frac{1+2n_B(\e)}{4\e^2+\w_n^2}\,, \label{eq:Pi_int}
\end{align} 
where $\e_k^2=\b k^2+m_T^2$, $n_B(\e)=1/[\exp(\e/T)-1]$ is the Bose-Einstein distribution,  
$K_d=\tfrac{1}{(2\pi)^d}\int\!\d\W_{d-1} = 2^{1-d}\pi^{-d/2}/\G(d/2)$
is the normalized area of the unit sphere $S_{d-1}$, and $\G(z)$ is the Gamma function.

\subsection{Asymptotics}
\begin{figure}   
\centering
\includegraphics[scale=.63]{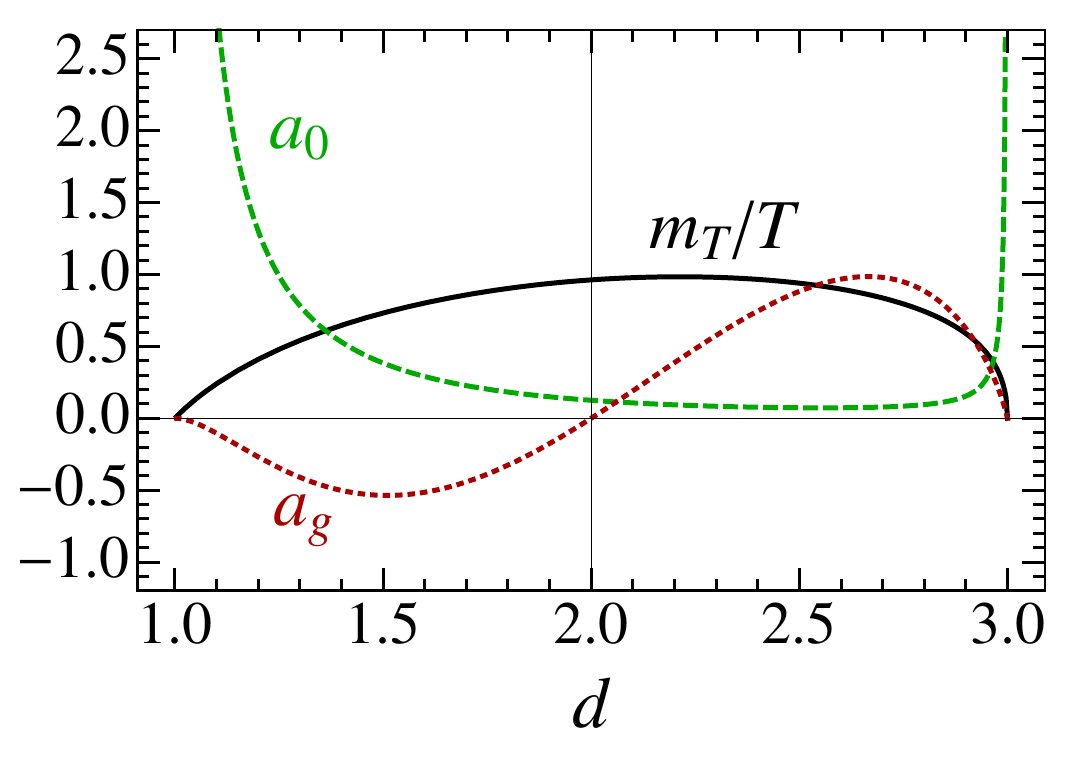}     
\caption{\label{fig:m} 
Thermal mass $m_T/T=\Theta_d$, as well as expansion coefficients $a_0,a_g$ appearing 
in $\chi_s(|k|\gg T)$ as a function of the spatial dimension $d$. Interestingly, $a_g$ vanishes in $2+1$D.
} 
\end{figure}

We find the following $\w_n\gg T$ expansion for $\Pi^{\rm E}$:
\begin{multline} \label{eq:pi-exp}
  \Pi^{\rm E}(i\w_n,\b 0)= \frac{a_0}{\w_n^{3-d}} \left(1 + \frac{1}{\w_n^{d-1}}\bigg[-\frac{(d+1)\G(d/2)\G(-(d+1)/2)m_T^{d-1}}{4\sqrt\pi} 
+ \int_{m_T}^\infty \!\!\d\e (\e^2-m_T^2)^{(d-2)/2}2n_B(\e) \bigg] \right. \\
\left. +a_g\frac{T^2}{\w_n^{2}}  - a_T \frac{T^{d+1}}{\w_n^{d+1}}+\dotsb\right)\,,
\end{multline}
where the dimensionless coefficients $a_\#$ are given by:  
\begin{align}
  a_{0} &= - \frac{K_d\pi\sec(d\pi/2)}{2^d}\,; \\
  a_{g} &= -\frac{K_d\Theta_d^2}{a_0} \frac{(d-2)\pi\sec(d\pi/2)}{2^{d-1} } \,; \label{eq:ag} \\
  a_{T} &= \frac{K_d\Theta_d^{d+1}}{a_0} \left\{ -\frac{\G(d/2)\G(-(d+1)/2)}{\sqrt\pi} +8\int_1^\infty\!\! \d\varepsilon
(\varepsilon^2-1)^{(d-2)/2}\varepsilon^2 n_B(m_T\varepsilon) \right\} \,.
\end{align}
$\Theta_d,a_0,a_g$ are plotted in \rfig{m} for $1<d<3$. The coefficient of the term arising
from the stress tensor, $a_T$, is positive for $1<d<3$.  
We find that the coefficient of $1/\w_n^{d-1}$ in \req{pi-exp} vanishes exactly 
once the value of the thermal mass $m_T$ is used (see Sec.\ts\ref{sec:therm_mass}).
This is as expected since the QC point does not have 
a O$(N)$-invariant scalar with scaling dimension $d-1$, and thus a $1/\w_n^{d-1}$ term in \req{pi-exp} would be at odds with the 
$\og\og$ OPE. An analogous cancellation was found\cite{ope1} to occur for the conductivity of the O$(N)$ model.  
\req{pi-exp} then leads to an asymptotic expansion of $\chi_s^{\rm E}$:
\begin{align} \label{eq:chi_s-asym}
  \chi_s^{\rm E}(k) = -\frac{N}{a_{0}} |k|^{4-D} \left( 1-a_{g}\frac{T^2}{|k|^2} +a_{T} \frac{T^D}{|k|^D} +\dotsb \right)\,,
\end{align}
where we have reinstated the the full dependence on $|k|^2=\w_n^2+\b k^2$. 
In two spatial dimensions the above simplifies to:
\begin{align}
  \chi_s^{\rm E}(k)=-8N|k|^3\left( 1 +\frac{2^8\zeta(3)}{5\pi} \frac{T^3}{|k|^3} +\dotsb \right)\,.
\end{align}
Interestingly, from \req{ag} we note that   
$a_g$ vanishes exactly in $d=2$ spatial dimensions (but not when $d\neq 2$). This stems from the fact that the OPE coefficient of
$\og$ in the $\og\og$ OPE, $\mc C_{ggg}$, vanishes exactly in $d=2$ and $N=\infty$. This was previously noted 
in Ref.\ts\onlinecite{petkou98}, where the expansion \req{chi_s-asym} was also given. 
As mentioned in the main body, we do not expect that $\mc C_{ggg}$ vanishes at finite $N$. 
In agreement with this expectation, the OPE coefficient in the Ising case ($N\!=\!1)$ was recently  
computed\cite{caselle15}
by means of Monte Carlo simulations and found to be finite.
It would be interesting to compare this new result with conformal bootstrap or with a $1/N$ expansion on the 
field theory side.  

\subsection{Thermal mass}\label{sec:therm_mass}
Interestingly, by imposing the vanishing of forbidden terms in the asymptotics of a 2-point function such as $\chi_s$,
we can \emph{determine} the value of the thermal mass, $m_T=\Theta_d T$.
Indeed, reverting back to the integration variable $\b k$ in \req{pi-exp}, we find that setting $[\cdots]=0$ 
leads to the following integral equation for $\Theta_d$:
\begin{align}
-\Theta_d T^{d-1} X_{d+1} + \int\frac{\d^d \b k}{(2\pi)^d} \frac{n_B(\e_{k})}{\e_{k}} =0\,,
\end{align}
where $\e_{k}=\sqrt{\b k^2+(\Theta_d T)^2}$, and we introduced the dimensionless
constant $X_{d+1}=2\G((3-d)/2)/[(4\pi)^{(d+1)/2}(d-1)]$. 
This agrees exactly with the equation obtained by requiring that
$\varphi_a\varphi_a=1$ in the non-linear sigma model\cite{csy,book}. 
In $D=3$, this equation can be exactly solved\cite{csy}: $\Theta_d=2\ln\varphi$, where
$\varphi=(1+\sqrt{5})/2$ is the golden ratio.     

\subsection{Response function and sum rule}  
The scalar response function is defined as follows:
\begin{align}
  \mc R_s(\w,\b k) = \frac{\chi_s^R(\w,\b k)}{i\w-0^+}\,,
\end{align}
where $\chi_s$ is the susceptibility. 
We shall set the spatial momentum $\b k$ to zero for simplicity.
Our goal is to explicitly verify the sum rule for $\mc R_s$.
Using the general result given in the main text, \req{sr-R}, we have at $N=\infty$:
\begin{align} 
  \int_0^\infty \d\w \re\left[\mc R_s(\w)-\mc R_s^{T=0}(\w) \right]=0\,, \label{eq:phi2-sr} 
\end{align}
which is precisely \req{chi_s-sr}. Using the asymptotics obtained above, we find that
$\mc R_s^{T=0}(\w)=(N/a_0)(\w/i)^{3-D}$, which becomes frequency independent for $D=3$.  
After subtracting this leading asymptotic behavior, the integrand decays sufficiently fast for the sum rule to be well-defined.
Indeed, using \req{chi_s-asym}, we have for $\w\gg T$: 
\begin{align}
  \mc R_s(\w)-\mc R_s^{T=0}(\w)=-\frac{(a_g N/a_0)T^2}{(-i\w)^{D-1}}+\dotsb\,,  
\end{align}
which comes from the relevant scalar $\og$, with scaling dimension $\De_g^{N=\infty}=2$ for $2<D<4$. 

\begin{figure}   
\centering
\includegraphics[scale=.77]{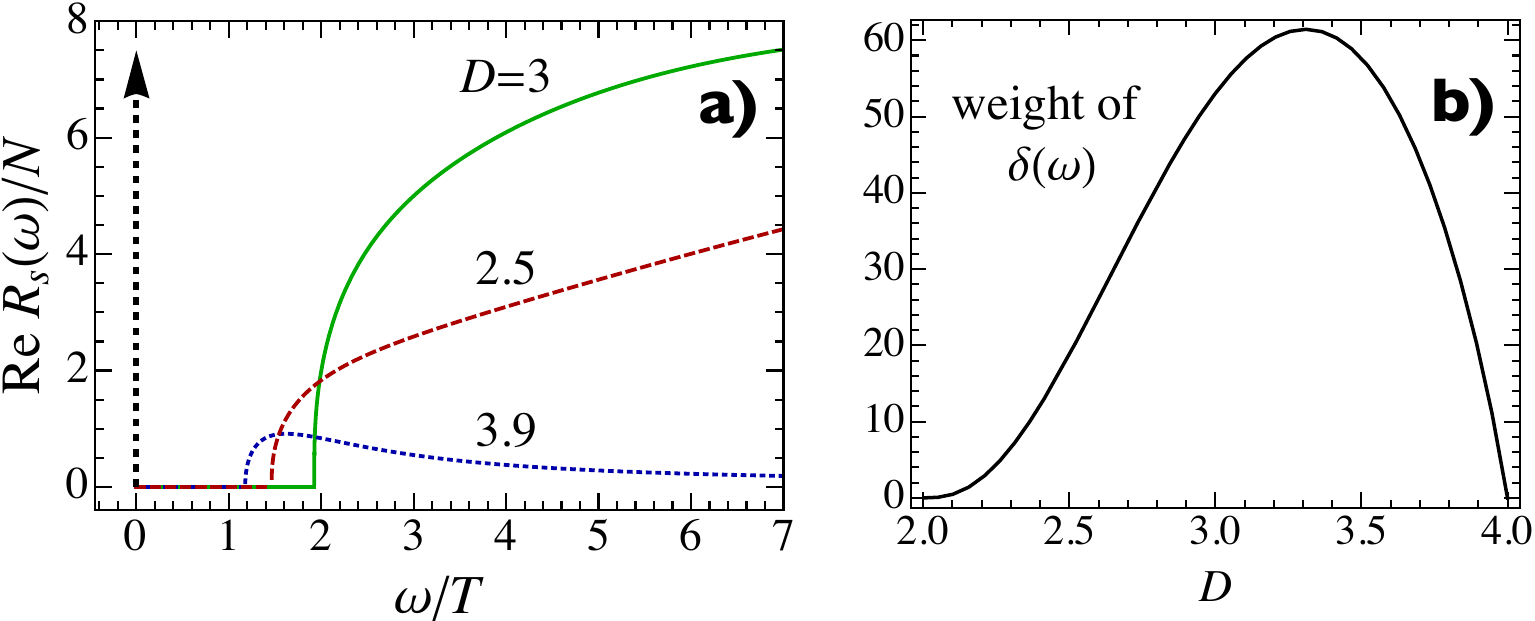}     
\caption{\label{fig:R_s} a) Scalar response of the O$(N)$ model ($N=\infty$) 
in various spacetime dimensions $D$. The dashed vertical arrow represents a delta function $\de(\w)$.
b) Weight of this delta function versus $D$.}  
\end{figure}

We now provide some details regarding the numerical verification of the sum rule. 
In general, the real part of the response is given by:
\begin{align} \label{eq:chi_s-largeN}
  \re\mc R_s(\w)=-\pi\chi_s^R(0)\de(\w) + \frac{1}{\w}\im\chi_s^R(\w) \,.
\end{align}
At $N=\infty$, we have\cite{csy,book} $\chi_s^R(\w)=-N/\Pi^R(\w)$, so that the spectral function for $\chi_s^R$ becomes
\begin{align} 
  \im\chi_s^R(\w)=N \frac{\im \Pi^R(\w)}{\left|\Pi^R(\w)\right|^2}\,,
\end{align}
where $|\Pi^R(\w)|$ is the complex norm. The imaginary and real parts of $\Pi^R(\w)$ at $\w>0$ 
can be obtained from \req{Pi_int}:  
\begin{align}
  \im\Pi^R(\w) &=\Theta(\w-2m_T) \frac{\pi K_d}{4\w} \left[(\w/2)^2-m_T^2\right]^{(d-2)/2}[1+2n_B(\w/2)]\,; \\
  \re\Pi^R(\w) &= K_d \, \mc P\!\!\int_{m_T}^\infty \d\e (\e^2-m_T^2)^{(d-2)/2} \frac{1+2n_B(\e)}{4\e^2-\w^2}\,,
\end{align}
% \begin{align}
%   \im\Pi^R(\w) &=\Theta(\w-2m_T) \frac{1+2n_B(\w/2)}{8\w} \\
%   \re\Pi^R(\w) &=\Theta(2m_T-\w)\frac{\tanh\inv(\w/2m_T)}{4\pi\w} + \Theta(\w-2m_T)\frac{\tanh\inv(2m_T/\w)}{4\pi\w} +
%   \mc P\!\!\int_{m_T}^\infty \frac{\d\e}{\pi}\frac{n_B(\e)}{4\e^2-\w^2}
% \end{align}
where $\Theta(x)$ is the step function, and $\mc P\!\!\int$ denotes the integral's principal value. 
We note that $\im\chi_s^R(\w)$ and $\re\mc R_s(\w)$ vanish for $0<\w<2m_T$, where $m_T$
is the ``thermal mass'' of the quasiparticles at $N=\infty$, \req{propagator}. This hard gap behavior is an artefact of the 
$N=\infty$ limit.
We further note that $\re\Pi^R$ has a logarithmic divergence at the threshold $\w=2m_T$. The resulting numerical plots for $\re\mc R_s(\w)$
are given in \rfig{R_s}a. Ref.\ts\onlinecite{will-mit} has computed $\im\chi_s(\w,\b k)$
at frequency and momentum $\b k$, where additional subtleties in the integrals emerge. The calculations of 
Ref.\ts\onlinecite{will-mit} can be used to verify the momentum-dependent sum rules, a task we leave for future investigation.
We finally note that $\re\mc R_s$ has a delta function at $\w=0$, \req{chi_s-largeN}, the weight of which is given by 
\begin{align}
  -\pi\chi_s^R(0)= \frac{N\pi}{\Pi^R(0)}\,,
\end{align}
which is plotted in \rfig{R_s}b as a function of $D$. For $D=3$, we obtain $N\times  53.04194$. 

The integral for the sum rule \req{phi2-sr} can be divided into three parts: the $\w=0$ delta function, the interval $(0,2m_T)$   
where $\re\mc R_s$ vanishes, and $(2m_T,+\infty)$: 
\begin{align}
  \mc I = -\frac{\pi}{2}\chi_s^R(0) + \int_0^{2m_T}\!\!\d\w \re\left[0-\mc R_s^{T=0}(\w)\right] 
  + \int_{2m_T}^\infty\!\!\d\w \re\left[\mc R_s(\w)-\mc R_s^{T=0}(\w)\right]\,.
\end{align}
For $D=3$, we numerically find:
\begin{align}
   -\frac{\pi}{2}\chi_s^R(0) &= +26.520972\,; \nn
 -\int_0^{2m_T}\!\!\d\w \re\mc R_s^{T=0}(\w) &= -15.398778\,; \nn
  \int_{2m_T}^\infty\!\!\d\w \re\left[\mc R_s(\w)-\mc R_s^{T=0}(\w)\right] &= -11.122193\,.
\end{align}
The terms sum to $\mc I=1.9\times 10^{-7}$. We can attribute the deviation from zero to the numerical uncertainty
in our evaluation of the integrals. (We emphasize that care must be used to 
treat the logarithmic vanishing of $\re\mc R_s$ at the threshold $\w=2m_T$.) The fact that three numbers of order $10^1$
cancel to within $10^{-7}$ provides an excellent check of the sum rule. We have also verified the sum rule
for other spacetime dimensions $D$, and have found that it is respected, in accordance with the proof given in the main body.   


%\bibliography{qc_dynamics}{}   
\end{document}